# Teachers of bachelors' lab courses collaborating to promote open inquiry: a case study

Lesley G.A. de Putter [a]* and Marloes M.H.G. Hendrickx[a]

[a]APSE- Eindhoven School of Education, Eindhoven University of Technology, Eindhoven, The Netherlands

*Lesley G.A. de Putter-Smits. L.G.A.d.Putter@tue.nl. +31402475936. Cascade 3.14. PObox 513. 5600 MB Eindhoven. The Netherlands

L.G.A. (Lesley) de Putter, ORCID: 0000-0003-1507-4785, is an assistant professor and teacher educator. Her research interests are STEM education, teacher education and continuing teacher professional development. A particular focus of her research is on engaging science learning environments. Science education is evolving towards student-centred learning, learning through scientific inquiry, within the context of social scientific issues.
M.M.H.G. (Marloes) Hendrickx. ORCID: 0000-0002-1290-3886 is an educational researcher and consultant. Her main research area is social interaction and relationships among teachers, among students, and between teachers and students. Her research particularly focuses on how collaboration can be supported and stimulated in impactful (STEM) education.

# Teachers of bachelors' lab courses collaborating to promote open inquiry: a case study

A group of university bachelors' teachers of open inquiry lab courses in science subjects collaborated in a Networked Faculty Learning Community (NFLC) with the final goal to share their course materials publicly. In this paper we describe their collaborative process, communication on the topic and their road to a common definition of open inquiry as a didactical method. The study emerged organically during the collaboration of this NFLC as the group increasingly recognized the value of documenting their endeavours, allowing others to learn and grow as they did. We can conclude that learning takes place in such a community, even though professional development wat not the NFLC's intention.

**Introduction**

Open inquiry as a didactical form in lab work for undergraduate students has received more attention recently, since it is a powerful means to teach students critical thinking within their discipline (Holmes et al. 2015). Bachelors' students across the science disciplines indicated that they learned not only how research was conducted, but also gained conceptual knowledge while working according to open inquiry (Holmes et al. 2017). Higher education teachers introducing open inquiry to their lab courses need support to implement such an innovation (Day et al. 2022, Talafian et al. 2025). A means to obtain this support is to have a group of peers wanting to learn about the same topic in a community of practice (COP) (e.g. Lave and Wenger 1991).

Open inquiry in higher education tends to be described per subject, such as for learners of physics (Walsh et al. 2022). There however undeniably are communalities in working according to this approach, since it rests on the research cycle scientists use. In this paper we study a group of university teachers that found each other through their common desire to share their experiences on open inquiry in their lab course. Five teachers from bachelor courses in three different departments formed this group to discuss how they could publish their open inquiry course materials such that other teachers were able to use their materials. They had obtained a higher education teacher grant to buy the time to work on this goal. In such a community learning is expected to take place with possible learning needs arising (Stevens et al. 2024). This case study describes the group's learning and their experiences while working on online shareable course materials for open inquiry lab work in higher education.

**Theory**

*Teachers working and learning together in a group*

Various terms have been used to describe groups of collaborating teachers, each reflecting different aspects of the group's structure and goals. The collaborating teachers used the term community of practice and referred to all the work together as 'work in the COP'. One of the teachers offered the term since they considered the definition of Lave and Wenger (1991) and applied that to their purpose: A group of teachers that all come from a different faculty or university, joined in the idea to strengthen their own courses that have open inquiry as a didactical form in common by exchanging teaching practices. Newer definitions might suggest the group can be seen as a Networked Professional Learning Community (NPLC) since not all participants are employed by the same university as is defined for a COP (Stoll and Louis 2007). The goal of the collaborating teachers in this study was however to also disseminate their way of working to other university teachers, inspiring and informing them on the use of open inquiry in bachelor's lab courses. The group's intended way of working comes closest to a Faculty Online Learning Community (FOLC) as first described by Dancy at al (2019) or a Virtual Learning Environment Faculty as described by O'Toole (2019). The group in this paper, however, also has the intention to meet in person, which is outside the definition of FOLC, whereas the Faculty Learning Community (Cox 2004) does not allow for people working together from different institutions. We therefore propose to see the group in this study as a Networked Faculty Learning Community (NFLC). A definition for the group, following Stoll and Louis (2007), can be: an inclusive group of people from different educational institutes, motivated by a shared learning vision, who support and work with each other mainly online, finding ways, inside and outside their immediate community, to enquire on their practice and together learn new and better approaches that will enhance all students' learning, with the intention to inspire and inform others of their vision.

Teacher learning and professional development in a group is known to take place when teachers are allowed to construct the learning setting and topic to be learned themselves (e.g. Garet et al. 2001). The agency the teachers experience on their own learning can motivate teachers to actually change their teaching practice (Van Veen et al. 2010; Cox 2004). The NFLC in this paper chose to work together to produce shareable course materials. Their plan was to discuss how open inquiry was situated in each course so the publication of the course materials would be in sync. The NFLC were supported by a grant, so they could afford to spend time on working in the group. The teachers thus had all opportunity to experience complete ownership of their working together, to asking each other questions and look for information as desired. In all, this NFLC has all the ideal characteristics for teacher professional development as described in a recent review on the topic of teacher professional learning (Stevens et al. 2024). The review describes three configurations of teacher professional learning groups, each differing from one another in 10 different characteristics, such as required expertise, format, support and intensity of the professional development. The NFLC in this study can be viewed as the configuration 'stimulating innovations', with nine of the 10 characteristics clearly present. The only difference the group has in comparison to the configurations Stevens et al. (2024) found was the impact category, that for the NFLC is not just the own organisation but all teachers of science lab courses.

The culture of the workplace influences a teacher's way of thinking about learning as well as the teaching behaviour (Lindblom-Ylänne 2006). This in turn influences the conversations about teaching in the NFLC, since the language they use for teaching and the teacher roles they have accepted as standard will differ for each individual. The teachers in the NFLC come from different departments, with different context and culture -even if from the same university-, social environment, expectations, with each teacher showing the expected behaviour that they derived from their working environment and their personal preferences (Prosser and Trigwell 1997). Struggles in finding a common language for describing the didactical form open inquiry can thus be expected. The discussions on teaching however also create an opportunity for teacher learning.

### *Open inquiry in lab work work*

The NFLC in this study shared the idea that they use open inquiry in their lab course. The term open inquiry has been around long enough for most teachers to have some idea of what it entails and for researchers to explore the possibilities of the use of this didactical form (Hofstein and Lunetta 2004). Learning through inquiry in education refers to using the steps of the research cycle, from research question and hypothesis to using literature and/or experiments to come to results and draw conclusions (Suchman 1965). The definition of open inquiry vented by one of the teachers at the start of the NFLC in this study was that of Tamir: '[…] the highest level of inquiry, the students have to do everything by themselves, beginning with problem formulation and ending with drawing conclusions.' (Tamir 1991 p16). Open-to-closed-inquiry then refers to the degrees of freedom the learner has. Closed-inquiries include a full description of all steps of the research cycle the learner should follow to prove pre-drawn conclusions on a given experiment. Open inquiries can for instance have no descriptions of the question or how to come to results and conclusions. These open inquiries have limitations to what apparatus, time and information might be available to the learner, but the steps of inquiry are not defined.

Open inquiry as a didactical form introduces students to the scientific method in an experiential fashion. This would align learning and assessment activities with intended learning outcomes (Biggs 1996) in bachelor's practical work. The overarching learning goal in natural science labs is to do empirical research and thus construct knowledge through the scientific method (Kozminski et al. 2014). Closed or structured inquiries have little in common with learning to work as a scientist (Holmes and Wieman 2016). Open inquiries allow the students to come up with a question or problem to solve, formulate hypotheses, design the experiment, and interpret their data to formulate conclusions. The use of open inquiry in undergraduate lab work gives students agency that trains their critical thinking (Ansell and Selen 2016) and motivates students to learn (Holmes and Wieman 2016). Students' conceptual learning is not hindered by using open inquiry as a teaching method, provided students are aware of the objective of the lab work (Holmes et al. 2017, Kirschner et al. 1993).

Although the general idea of open inquiry is clear, the actual application of open inquiry in a specific lab course will differ, from the institution involved to the science topic studied, the undergraduate year, available equipment, time, up to how the course is assessed. A common definition is therefore cumbersome and bachelor's level teachers trying to elicit how to change their course into open inquiry can have a hard time finding out how (Bauer and Emden 2021). The teachers in this study stem from different disciplines and departments, so although they consider open inquiry is their didactical form, the actual execution will be different.

This case study aims to describe the group's learning and experiences as they collaborated to create publicly available materials for open inquiry lab work in bachelors courses. Central to this process is how the group negotiated a shared understanding of open inquiry. Therefore, the research question is: 'How does a group of teachers from different universities and departments and disciplines come to define open inquiry in bachelor's practical work based on their own courses and their interpretation of literature?' we defined two sub-questions:

1. How can the process and reasonings be described when a group of teachers sets out to define open inquiry in bachelor's practical work based on their own courses and their interpretation of literature?

2. How can the content steps toward a final definition of open inquiry in bachelor's practical work based on their own courses and their interpretation of literature be described?

*Backdrop to the study*

The study is situated in the Netherlands. The NFLC consisted of five teachers in higher education that came from five different departments from three different universities, teaching different science subjects to different years of bachelor's students, and one educational researcher as mediator in the meetings. Together, they applied for and received a so-called SURF grant (SURF is the ICT cooperative of Dutch education and research institutions) to be able to buy time from their departments to work together on the topic that brought them together: the open inquiry teaching method in bachelor's practical work. The grant facilitated time for the meetings with the goal to create shareable online versions of their course materials in order to provide other teachers with an example on how to work according to open inquiry. The SURF organization offers a free digital platform to share these course materials with other teachers. The teachers come from the university departments A to E (in Table 1). They each are the responsible teacher for the course. Some have teaching assistants or lab technicians to help out, some share the course content with another teacher (not in the NFLC). In two cases there is a restriction in lab time available for the course, which means students have to be ready to do their experiments at a certain time and finish within a given time. The other courses have the lab available to the students during the course's normal time, directed by the department's schedule.

Table 1. Overview of participants and the courses they are responsible for

| Department | Teacher | Subject | BSc Year | Open to | Credits |
|---|---|---|---|---|---|
| A | Blake | Science | 2/3 | Elective, yr 2 or 3 | 6 |
| B | Alex | Physics | 3 | Compulsory | 6 |
| C | Sam | Molecular neuro-biology | 3 | Elective, all in yr 3 | 6 |
| D | Remi | Advanced genomics | 3 | Elective, all in yr 3 | 6 |
| E | Parker | Maker Course | 3 | All universities | 30 |

The initiation phase of the project was during the COVID-19 pandemic, allowing online meetings only. It was decided to use the form of an online NFLC with site and course visits as soon as this became possible. The teachers had opted for an NFLC to facilitate learning from each other by presenting the ideas behind their lab courses and discussing these with their peers.

In their online meetings they discussed their ideas about teaching, specifically on open inquiry with one another, experiencing differences in meaning or operationalization of words. They compared notes on assessment, student guidance, handling problematic situations with students ('the guy kept on showing up drunk still, from last night's bender') and involved each other in the small redesigns they made in the two years the NFLC was running. A researcher was involved in the NFLC from the start and included in all email conversations and meetings.

**Method and Materials**

*Research set-up*

Ethics approval for the research project was obtained from the researcher's institute under number ERB2022ESOE13. All participants gave their full consent to use all data. The teachers reviewed and approved the present paper, but they were not involved in performing the analyses or describing the results.

There were 30 meetings in total, starting from the first idea to start an NFLC (Spring 2022) and apply for a grant to content filled meetings on the courses, visit to running courses, the assessments, and the desired consensus on open inquiry to share with others, to the final meetings where the topic was the sharing of course materials online (September 2024). The 14 meetings dealing with content up to the final session on the shared definition of open inquiry are part of this paper. The first meetings were mainly organizational and financially oriented, and the last meetings were focused on the intricacies of the online software where the course materials should be published in.

From these 14 of 30 meetings (from October 2022 until June 2023) the researcher collected all emails and PowerPoint presentations, the meeting notes (made by researcher, or together with the NFLC on a Miro Board online), and iterations of documents annotated by the teachers with concept definitions on open inquiry. The course materials such as the student manual with the learning goals, course description, assessment plan and so on that the teachers shared on a common network drive were also used.

*Analysis*

All data were sorted in chronological order. First the data set was analysed for the process and reasonings (question 1); next, for the steps leading to the final definition. Since the researcher was involved throughout the whole project, an independent researcher was attracted who had no experience with the group or the data, so unbiased analysis could be assured.

The data were qualitative, and not purposefully collected. Therefore bottom-up coding was used in the first analysis step by indicating text fragments that pointed towards process or reasoning of each teacher. The researchers both identified fragments separately in the total data set. They then compared fragments identified and discussed until full agreement on the fragments to be included was reached. Next the researchers discussed the fragments and concluded that they could be categorized as: teacher brings information on open inquiry elements to the group; teacher takes information on open inquiry away from the group to use; teacher endeavours to reach consensus on open inquiry. The researchers each coded the same halve of the fragments and then compared results. Differences were discussed until agreement was reached. The first researcher then coded the remainder of the fragments. The second researcher audited the coding and where necessary coding was changed. Example fragments of each category are provided in Table 2.

Table 2. Example fragments of the three categories identified.

| Category | Description | Example quote |
| --- | --- | --- |
| Brings information | In the fragment the teacher offers information helpful to open inquiry stemming from their course or from literature, from their faculty or hearsay | 'I just came across this article on generating creative ideas in biology inspired design. I think it is right up your street but among other parts I found the ideas on creativity also very interesting!' |
| Takes information | In the fragment the teacher indicated to use or wants to have information for their own course design or their own learning on open inquiry. | '.. at some point in time I would like to discuss the way we can deal with the way students feel that they are getting conflicting advice from different teachers/TA's.' |
| Tries for consensus | In the fragment the teacher suggests a formulation combining information from more than the own course or otherwise suggests a means to or agreement with a common definition of open inquiry | 'This is indeed my timeline!' |

Keeping the data in chronological order and looking at the occurrence of the categories over time, there seemed to be roughly three phases in the process (see figure 1). A start phase from September to December 2022 where bringing information more than taking and consensus take place; a change-phase in January and February 2023 where the bringing lessens and taking and consensus start occurring (more); a negotiation phase from March to May 2023 where bringing and consensus are most notable; and a final phase where consensus is predominant (June 2023). We will describe the phases of the process and the changes in the fragments per phase in the results.

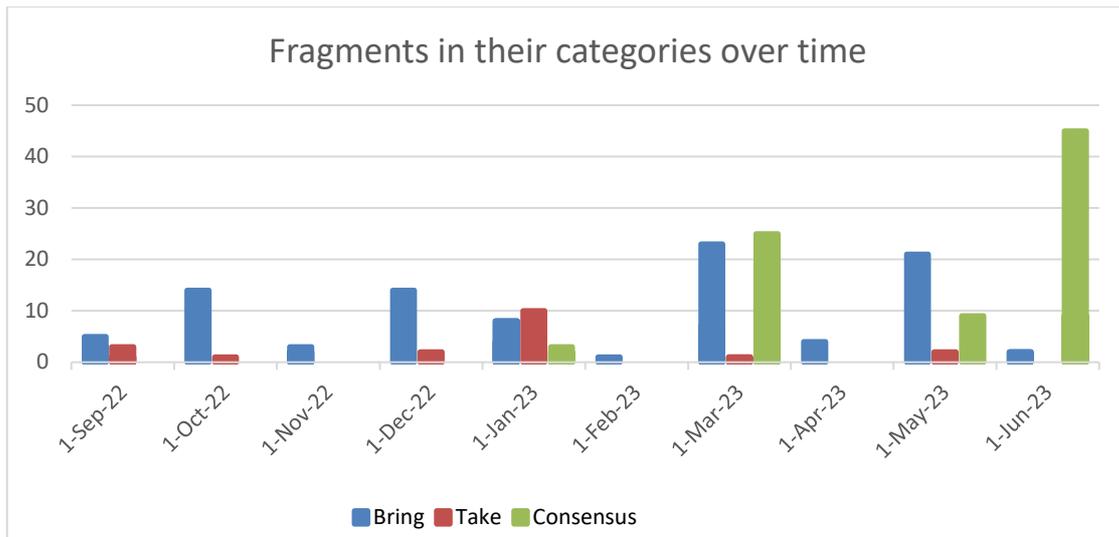

Figure 1. An overview of the number of fragments indicating bringing, taking or consensus making over time.

To answer sub-question 2 the data were analysed for additions, suggestions and arguments that were aimed at forming a common definition of open inquiry. This process was performed foremost with the team of teachers. The second researcher audited the data and different concepts of the definition of open inquiry to ensure all teachers' additions, opinions, and suggestions were taken into account.

**Results**

*Phase one*
In the first three months the teachers mainly explain how they designed their open inquiry lab course, what materials they use, the didactical knowledge they have, publications they use to inform their teaching, and experiences with running the course. During the discourse teachers ask explanatory questions, with some teachers indicating being inspired by the other teacher's work and wanting to incorporate this in the own course. The phase is a bringing and taking phase, with the 'bringing' being the most prominent.

    Alex: 'I don't use that, or maybe I do, I use the steps of research more.' Sam: 'I probably do that too, although I did not see a model before. They just work like a researcher.' (-researcher notes)

Blake: 'My bosses suggest that saying explicitly that we're training empirical research skills (and using the phrase 'undergraduate research') aligns better with the outcome-focused zeitgeist, as the "open inquiry" teaching method may not excite people as much as 'undergraduate research' (-summary ppt)

The teachers start describing what they learn or take away from the meetings towards the end of the first phase, as can be seen from the quote below:

Alex: 'What I distilled from our meeting: Do we actually do inquiry based learning? What do students learn in inquiry based learning compared to the learning goals the teachers have? Which things did students learn about doing (inquiry based) research? Which things would students use again in future research?' (-email exchange)

*Phase two*

After the third month the teachers realize that there are differences in approach, but agree that there seem to be common denominators allowing them to take up ideas from others to improve the own education more and more. They start to express their need for a definition of open inquiry lab teaching. During month four and five the bringing and taking almost even out and the consensus remarks start occurring. During this time the NFLC actively lobbies to have other lab course teachers joining the meetings, which occurred twice. The two visiting teachers were from different universities, looking to change their lab course didactic strategy.

The live meet-up where the NFLC could visit each others' courses and explain both visually and in words what their course was all about seems a pivoting point in understanding the similarities in their courses. Each teacher enthusiastically spoke about their course and in two cases the students were there to provide insights on their ideas about the way the lab course was run.

Sam: 'Students often come up with one main research questions and two sub-questions. The assessment, students receive an individual grade for academic attitude and laboratory work. The group as a whole is graded on their performance during journal club, the two lab meetings and the final presentation. When someone in a group really stands out during the final presentation for instance they can be given a higher, or lower, grade.' (-notes during live meet-up)

An example of a quote indicating a teacher taking information away to the own course during phase two is:

Parker: 'The bio-lab visit was inspiring, as it seems like a nice fit for some of our (future) students' (-open-ended questionnaire after live meet-up)

The NFLC discusses the definition of open inquiry through a PowerPoint that has been placed on an online Miro board. Sticky notes with comments are added indicating additions, rephrasing and remarks indicating the teacher does not feel something is included in their course, although not disagreeing it might be part of open inquiry, a mere: 'we don't do this'. -Parker. Mixed ideas also come up, trying to conciliate student agency and the need for scaffolding:

Remi: 'I try not to give them example reports to force them to come up with their own ideas' (sticky note on ppt)

The remarks teachers make change from phase to phase. In phase one the tone is questioning, careful and hesitant, using words such as: maybe and probably. In phase two the tone is more open and sharing, talking about the own course in detail and this being received with invitations to elaborate and suggestions to further the conversation:

Alex: Does not need to be the next one but at some point in time I would like to discuss the way we can deal with the way students feel that they are getting conflicting advice from different teachers/TA's. (-open-ended questionnaire after live meet-up)

The tone changes again in phase three, where the urge to have consensus and bring the common core to the public is fore fronted in the conversations.

*Phase 3*

From month six a lot of effort goes into reaching consensus which causes teachers to bring more information to clarify how they designed their teaching. The taking of information still takes place, although less prominent, because he focus shifts to reaching consensus on the common definition of open inquiry in lab work.

Definition 1:

'Open inquiry in lab courses is when students work like a scientist, using the experimental cycle, learn practical and academic skills and strengthen their scientific knowledge.' (meeting ppt)

After this first attempt, the NFLC starting discussing that such a general definition did not do justice to all the things open inquiry entails. They started creating a list of items that should become part of the definition, to be put in a sentence later. The items were categorized in two more attempts to place the items under a category that seemed correct to the NFLC. The items were discussed and rephrased during meetings in phase three. Definition two is provided in Table 3 below.

Table 3. The second definition of open inquiry of the NFLC (-meeting ppt)

| Students: | |
| --- | --- |
| Work in an authentic setting: | Real stakeholder/problem/object |
| | Research setting |
| Work autonomously from teachers | Self-regulated learning |
| | Collaboratively |
| | Peer feedback & intervision |
| | Scaffolding |
| Formulate their own question: | Research question and hypothesis |
| Design problem and possible solutions | |
| Look for information: | Experts |
| | Books |
| | Web |
| | Literature |
| Follow expert ways of working: | Steps of research |
| | Design process |
| Present their results to others: | Presentation |
|                                      | Report                             |
|                                      | Article                            |
| Show their mastery through reflection: | What they learned                |
|                                      | What their results mean to others  |

A session where each part was discussed and added onto the second definition gathered more detailing and categorizing. The final definition of open inquiry after the sensemaking process is a one-liner, followed by detailing of what each of the items in the one-liner actually means to them.

The final definition the NFLC agreed on is provided below in Table 4.

Table 4. The final definition of Open Inquiry the FOLC agreed on

| *'Open-inquiry in bachelor's lab courses requires students to complete an entire, authentic scientific research or design process, coached in their work by teachers.'* ||
|---|---|
| Students working through open inquiry: | The process coaching involves that: |
| (1) Work in an authentic setting, which implies:<br>   a. Real stakeholder or person/problem/object<br>   b. (Authentic) Research setting – work like a researcher<br>   c. Both failure and success are possible acceptable (learning) outcomes.<br>(2) Follow expert ways of working, such as the ways of a:<br>   a. Scientific researcher<br>   b. Designer<br>   c. Professional (company)<br>   d. Iterative steps of design- and research processes<br>   e. Includes a safe working environment or practice.<br>(3) Formulate a design problem or research question and hypothesis, which includes:<br>   a. Design problem and possible (multiple) solutions or plans<br>   b. (quantitative) Research question and hypothesis<br>   c. Testable hypothesis<br>   d. Formulate when their design or research is successful (or not) and how they know.<br>(4) Consciously and critically look for information from:<br>   a. Experts (teaching assistants, researchers, teachers, students from previous years, content people) | (1) Students are aware of the time-frame, in that they can:<br>   a. Keep track of time/planning<br>   b. Reflect on what they can achieve in given time<br>   c. Take availability of lab and materials into account.<br>(2) Students can complete the course's assessment, including:<br>   a. Presentations<br>   b. Half-way tests, reports, products<br>   c. Deliverables.<br>(3) Work autonomously from teachers, which means they work through:<br>   a. Self-regulated learning<br>   b. Collaboration<br>   c. Peer feedback & intervision<br>   d. Scaffolding. |

b. Books, the web, literature
   c. Other students
   d. Interpreting, questioning, reflecting on the information and advice from others
   e. Dealing with conflicting information and simplifications.
(5) Concisely (re)formulate design criteria or a hypothesis, which means:
   a. A description of an acceptable means to arrive at a quantifiable, measurable answer to the research question or design problem
   b. Returning to the description to reformulate in an iterative manner.
(6) (re) Design a prototype or set-up an experiment, which means:
   a. Make choices
   b. Check on safety
   c. Iterate.
(7) Construct a prototype and/or run the experiment, which includes:
   a. Repetition
   b. Quick test results,
   c. Calibrations,
   d. Feasibility,
   e. Uncertainty,
   f. Pilot studies
   g. Having alternative plans.
(8) Interpret results, these include:
   a. Relate to design and hypothesis
   b. Reliability, Error
   c. Reproducibility
   d. Validity
   e. Iterative -when possible in view of time, materials etc..
(9) Report on results in various ways:
   a. Present their results to others in:
      i. Presentation
      ii. Report
      iii. Article
   b. Defend results and choices and decisions made
   c. What the results mean to others, implications
   d. Successes and failures.
(10) Draw conclusions, where they
   a. Describe their conclusions to the research question
   b. What the conclusion means to others; implications.

**Discussion and Conclusions**

The current study followed a group of university teachers in their endeavours to promote open inquiry as a teaching method in lab work. The first research question on how the process and reasonings can be described when a group of teachers sets out to define open inquiry in bachelor's practical work based on their own courses and their interpretation of literature, was answered by describing the differences in tone and kinds of remarks that follow a pattern in time. The driving force of the community was disseminating a didactical form the teachers were all convinced of, not teacher learning. Comparing the categories we found (brings information, takes information, consensus) with the well-known interconnected model of professional growth (Clarke and Hollingsworth 2002) however, it can be argued that they can be roughly mapped on the model. The data categorized as teacher bringing information to the NFLC stem from the personal domain and the domain of practice which are difficult to distinguish in our data. The external domain is interaction with the NFLC itself and the data categorized as teacher taking information away is the domain of consequence in the model. Teacher learning was not the intention of the NFLC, but the discourse over time and the objective the NFLC had apparently shows a pattern of interactions that is similar to that described by the interconnected model of professional growth.

The reasonings, or talk pattern of the NFLC identified over time can be viewed from Mercer's categories of talk (Mercer 1996). The NFLC data show cumulative talk in the first phase of the process described here. The group shares information to construct this common knowledge on open inquiry. Naturally, the group ventures from cumulative to exploratory talk when the differences in the information each teacher offered were accepted and a second goal emerged: the need for a consensus on open inquiry in lab courses to share with all teachers. The group's shift toward exploratory talk reflects a move towards more meaningful collaboration. This form of talk is associated with increased professional learning outcomes, as it entails critical reflection, mutual challenge, and the co-construction of meaning.

The second research question was how the content steps toward a final definition of open inquiry in bachelor's practical work based on their own courses and their interpretation of literature can be described. The NFLC went from a statement style definition to a more and more detailed and comprehensive definition. The idea of a common definition grew from the desire to have a workable common definition to having design indicators for other teachers to design an open inquiry lab course with. Open inquiry as a didactical method has its advocates and its antagonists, where both groups indicate that it is not a good idea to have students jump-in the water without knowing how to swim (Hofstein and Lunetta 2004, Kirschner 1992). Definitions of open inquiry in research on lab courses are generally not provided. Each research states the focus of the research, such as teaching students critical thinking (e.g. Holmes et al. 2015) or the learning goals of the lab courses in question, for instance cookbook style experiments versus more open inquiry style (e.g. Blanchard et al. 2010) in which general circumstances are described. Inquiry learning or learning the nature of science papers and policies refer to a list of characteristics in such education (e.g. Brock and Park 2022, Kozminski et al. 2014), not perse specified to lab courses in higher education. The steps the NFLC took to reach a consensus on a definition that could describe all their lab courses at the same time are a rarity.

The content of the definition in this paper allow for two fundamentally different learning goals: namely a scientific open inquiry and a design open inquiry. All elements of the definition however hold, for both goals. Looking closely at the final definition they appear to cover all elements of the curricular spiderweb as presented by Van Den Akker et al. (2003), specified for the design of open inquiry lab courses. They represent empirical design criteria that have to be determined for each course to make the educational design work. The group, realizing what they had created, promoted dissemination of this model in an open online educational materials website and sharing this model with colleagues from higher and secondary science education.

The study described here was started ad-hoc, on request of the participants, to make the most out of all the 'amazing things' they were experiencing together. Had the research been planned in advance, a different set-up would have been chosen, to capture more of the teacher's thinking, motifs, beliefs and experienced learning. As it is, it is a case study describing the journey of a enthusiastic teachers into open inquiry as a teaching method for undergraduate science lab courses. The next step is to compare the student experiences from these courses to uncover how they view the open inquiry lab courses, and whether their experiences match the teachers' intentions.

**Acknowledgements**

We like to acknowledge all teachers in the NFLC for their enthusiasm, consent and cooperation in doing this research;

**Declaration of interest**

The group of teachers and the main researcher of this study received a grant to work together on creating online open teaching materials. The proposal for this grant included the desire to do research on both the process and the effectiveness of the materials. This desire was granted by the funding organization and each teachers' institute. The group was funded both by the grant and the own institute to afford every member the time to meet up and work on the materials to be shared online. To ensure the research would not be biased, an independent researcher was involved throughout the analysis. This researcher had no prior knowledge of the group and studied the anonymized data only.